\documentclass[12pt]{iopart}

\usepackage{graphicx}
\usepackage[square,sort,comma,numbers]{natbib}
\usepackage{xcolor}

\begin{document}
\title[]{Strain Induced Enhanced Photocatalytic Activities in Layered Two Dimensional C$_2$N/MoS$_2$ Heterostructure: A Meta-GGA Study}

\author{Soumendra Kumar Das$^1$, Lokanath Patra$^2$, Prasanjit Samal$^1$, Pratap K Sahoo$^{1}$}

\address{$^1$School of Physical Sciences, National Institute of Science Education and Research (NISER) Bhubaneswar, HBNI, Jatni, Khurda-752050, Odisha, India.}
\address{$^2$Department of Mechanical Engineering, University of California Santa Barbara, CA, 93106, USA. }
\ead{psamal@niser.ac.in, pratap.sahoo@niser.ac.in}
\vspace{10pt}

\begin{abstract}
  The improved photocatalytic water splitting using 2D materials has technological importance for economically viable renewable energy. The present study focuses on the effect of uniaxial, biaxial, and vertical strain on the energy gap and band edge positions of C$_2$N/MoS$_2$ van der Waals heterostructures through first-principles density functional theory using PBE and SCAN functionals. The calculations establish that SCAN functional provides comparatively much better results as compared to the PBE for the band gap and band alignment study. The heterostructure exhibits a type-II band alignment which is beneficial for the efficient separation of charge carriers. For a good photocatalyst, the band edge positions should straddle the water redox potentials. It is observed that for both compressive and tensile vertical strain, the water redox potential values lie within the valence band maximum (VBM) and conduction band minimum (CBM) of the heterostructure. On the other hand, for uniaxial and biaxial strain, the system can be used as a useful photocatalyst only for larger compressive strain, whereas for tensile strain, the energy gap between VBM and CBM keeps on decreasing and lie within the water oxidation/reduction potential. Our study also establishes that the meta-GGA SCAN functional shows similar results as compared to the computationally expensive hybrid HSE functionals. The present work can be extremely useful for experimentalists to design artificial heterostructure devices for better performance in photocatalytic water splitting.
\end{abstract}


\submitto{\TDM}

\section{Introduction}
The emergence of photocatalytic water splitting has been a successful technology to meet the demand for the energy crisis and environmental pollution created by our fast-growing economy. The development of high-performance photo-catalytic materials to create hydrogen by using solar energy has been a serious focus of research for many years ~\cite{Song2020, Takata2020}. The key factor for achieving highly efficient photocatalysts (PCs) is that the band gap should be larger than the water redox potentials. More specifically, the conduction band minimum (CBM) of the PCs should be above the H$^+$/H$_2$ potential and the valence band maximum (VBM) should be below H$_2$O/O$_2$ potential simultaneously, thus requiring a minimum band gap of 1.23 eV~\cite{liao2014design}. In addition, literature reports have established the importance of  co-catalysts for boosting the electron-hole separation and improving the reaction kinetics~\cite{zhang2020direct}. Under such circumstances, two-dimensional materials like graphene, hexagonal boron nitride (h-BN) mono layers, transition metal dichalcogenides (TMDCs), C$_3$N$_4$, C$_2$N, etc, have created a lot of interest, in meeting the demand, because of their novel electronic, thermal and optoelectronic properties. In particular, MoS$_2$ has a direct band gap (2 eV), and high carrier mobility in the form of a single monolayer, which makes it an important candidate for photocatalytic and photovoltaic applications~\cite{chhowalla2013chemistry, bonaccorso2015graphene}. Similarly, the porous C$_2$N monolayer is found to be a direct band gap semiconductor with a gap of 1.96 eV~\cite{mahmood2015nitrogenated}. Zhao et al. have adopted a 2D/2D polymeric Z-scheme heterostructure by using a pair of ultrathin g-C$_3$N$_4$ nanosheets in order to provide H$_2$- and O$_2$- evolving photocatalysts through the strategy of electrostatic self-assembly. Using Pt and Co(OH)$_2$ as co-catalysts, the heterostructure achieved a solar-to-hydrogen efficiency of 1.16 \% which originates due to the formation of direct Z-scheme charge transfer pathway through the interface between H$_2$- and O$_2$- evolving components ~\cite{Zhao2021}. It has been suggested that the use of C$_2$N and MoS$_2$ can be highly efficient for photocatalytic study and also can be complementary to the use of graphene and h-BN~\cite{mahmood2015cobalt}.  

Despite the extensive use of C$_2$N and MoS$_2$, there are some challenges as well for the application of these materials for photocatalytic study. The charge distribution of the valence band maximum (VBM) and conduction band minimum (CBM) states for these systems are not well separated in space resulting in reduced light absorbing efficiency because of the recombination of the photoinduced electrons and holes~\cite{ataca2012stable, mahmood2015nitrogenated}. Therefore, attempts have been made to use van der Waals heterostructures to fix the issues. The electronic properties of C$_2$N/InSe heterostructure are found to be greatly affected by vertical strain and electric field. Without any electric field, the heterostructure possesses a type-II band alignment with an indirect band gap of 1.34 eV at an equilibrium interlayer distance of 3.325 \AA. Application of an electric field or a change in interlayer distance results in a transition from type-II to type-I band alignment and indirect to direct band gap in this heterostructure~\cite{Pham2019}. 

Band gap and band offset engineering at C$_2$N/MSe$_2$ (M = Mo, W) interface have established that the heterostructure possesses a narrow indirect band gap with type-II band alignment which is favourable for the photogenerated electron-hole pairs. The application of vertical strain and electric field strongly modulate the magnitude of band gap values and band offsets, but the type-II band alignment nature remains preserved~\cite{Slassi2022}. Strain engineering plays a significant role in tuning the electronic properties and photocatalytic performance of 2D heterostructures. Wang et al. have studied the effect of strain on the electronic structure of C$_2$N/MTe (M= Ga, In) through DFT calculations which exhibit excellent optical properties with good structural stability. It was observed that for C$_2$N/GaTe heterostructure, the exciton-Bohr radius remains unaffected by the application of strain. On the other hand, compressive strain reduces the exciton-Bohr radius in  C$_2$N/InTe system. The power conversion efficiency shows an increase up to 22.1\% for  C$_2$N/GaTe with 4\% strain and 19.8\% for C$_2$N/GaTe heterostructure with 6\% strain ~\cite{Wang_new_2020}. Han et al. reported that the catalytic efficiency of C$_2$N/SiH heterojunction can be effectively adjusted by the application of -2$\%$ and +4$\%$ biaxial strain for the hydrogen evolution reaction (HER) and oxygen evolution reaction (OER), respectively~\cite{Han2022}. The Cs$_3$Bi$_2$I$_9$/C$_2$N heterostructure exhibits a charge distribution across the whole structure due to the difference in the work function between the two monolayers and a charge transfer at the interface due to the formation of an internal electric field~\cite{Wang2022}. The photocatalytic study in MoS$_2$/ZnO heterostructure shows an indirect band gap with type-II band alignment, a significant built-in potential of 7.42 eV, and a valence band offset of 1.23 eV across the interface. The photogenerated carriers are localized in different layers and can effectively generate hydrogen energy. In contrast, the MoSe$_2$/ZnO heterostructure possesses a type-I band alignment with a direct band gap of 1.80 eV, built-in potential around 3.64 eV, and a valence band offset of 0.34 eV~\cite{Sharma2022}.

Despite the extensive studies on C$_2$N based van der Waals heterostructures, significant progress has not been achieved both theoretically and experimentally. As a typical case, reports on the C$_2$N/MoS$_2$ heterostructures are very scarce and a systematic investigation of the physical properties of this system is still lacking. In this article, we report the effect of vertical, uniaxial, and biaxial strain on the photocatalytic water splitting performance of C$_2$N/MoS$_2$ van der Waals heterostructures through first-principles electronic structure calculations. 
 
\section{Results and Discussion}
The schematic of the crystal structure of C$_2$N monolayer, MoS$_2$ layer, and C$_2$N/MoS$_2$ heterostructure is illustrated in Fig. 1. In C$_2$N monolayer, twelve C-N bonds are alternately connected to the six C-C bonds in such a way that the periodic holes are present without any atom in the middle of the lattice. The in-plane bond length is estimated to be around 1.43 and 1.47 \AA ~for the C-C bond and 1.34 \AA ~for the C-N bond, respectively which is also close to the  previously reported values~\cite{Kishore2017}. The optimized lattice constant using PBE functional is estimated to be 8.33 \AA ~and 3.19 \AA ~for the C$_2$N and MoS$_2$ monolayer, respectively which is in reasonable agreement with the reported experimental and theoretical results~\cite{Wang2012, Mahmood2015}. The heterostructure is constructed by making a supercell of (3$\times$3$\times$1) for C$_2$N and (8$\times$8$\times$1) for MoS$_2$ layer, respectively. The corresponding lattice mismatch between the two monolayers is around 1.9$\%$ which is within the acceptable range. The heterostructure after ionic relaxation results in a corrugated structure (Fig. 1d), very similar to silicene. The initial van der Waals gap (3.5 \AA) between the layers reduces to 3.373 \AA ~with an optimized lattice constant of 25.47 \AA ~(Fig. 1c, d).

\begin{figure}[!htb]
\centering
\includegraphics[width=1.0\textwidth]{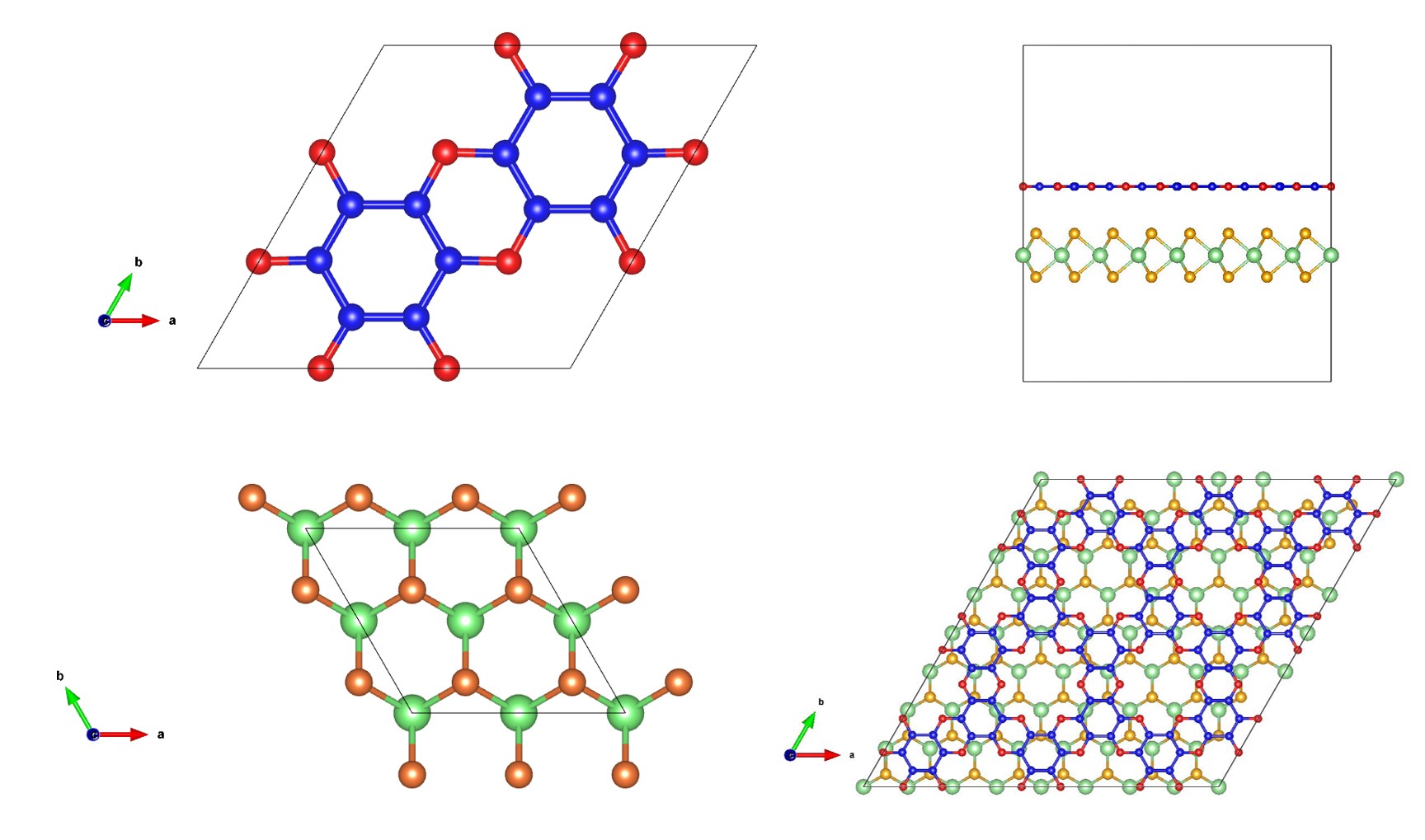}
\caption{Schematic representation of the top view of isolated (a) C$_2$N monolayer, (b)  MoS$_2$ monolayer, (c) side view and  (d) top view of C$_2$N/MoS$_2$ heterostructure. The schematic symbols of the C, N, Mo, and S atoms are indicated as blue, red, green, and brown colors, respectively.}
\label{Schematic}
\end{figure}

The calculated electronic structure of free-standing C$_2$N monolayer, MoS$_2$ using the PBE functional are given in Fig. 2a,b. The band dispersion for both the material indicates the presence of valence band maximum  (VBM) and conduction band minimum (CBM) at the same \textbf{k} value in the Brillouin zone. In addition, for both structures, the VBM remains close to the Fermi level as compared to the CBM. This confirms that both  C$_2$N and MoS$_2$  monolayers are p-type direct band gap semiconductors and the calculated energy gap values are estimated to be 1.735 eV and 1.645 eV, respectively.  It is interesting to note that the energy band dispersion in C$_2$N monolayer is relatively flat where as MoS$_2$ shows highly dispersive bands both in VB and CB regions. Hence, it is expected that the MoS$_2$ system will have a relatively low electron and hole effective mass along the high symmetry path ($\Gamma$-M-K-$\Gamma$) as compared to that of C$_2$N monolayer. Therefore, the electron and hole mobilities of MoS$_2$ would be larger than that of the C$_2$N monolayer, since the effective mass is inversely proportional to the carrier mobility. Figure 2e illustrates the electronic structure of C$_2$N/MoS$_2$ heterostructure which preserves the direct band gap behaviour of the isolated monolayers. The VBM is situated close to the Fermi level suggesting that the charge carriers are of p-type. However, the value of the energy gap is comparatively reduced and becomes 1.353 eV. We note that the ideal band gap for a semiconductor to use more visible light is around 1.5 eV~\cite{chowdhury2017monolayer}.

\begin{figure}[!htb]
\centering
\includegraphics[width=1.0\textwidth]{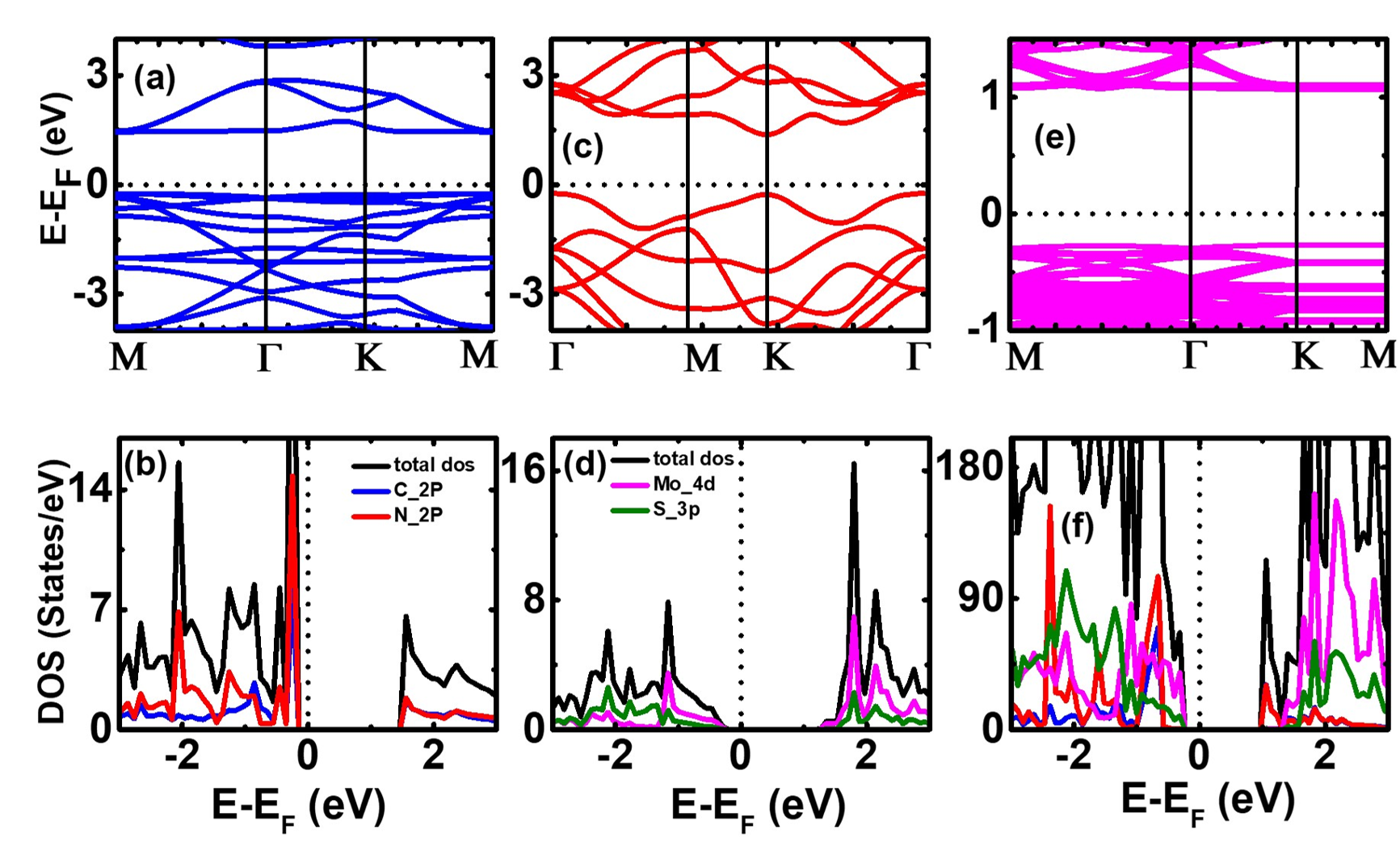}
\caption{Electronic band structure of (a) C$_2$N, (b)  MoS$_2$  monolayer, (c)  C$_2$N/MoS$_2$ heterostructure, (d-f) The total and projected density of states for the corresponding system calculated using PBE functional. The dashed line in each figure indicates the Fermi level.}
\label{Schematic}
\end{figure}

It is well known that the PBE functional severely underestimates the energy gap of the system. Again calculations involving the hybrid density functional like HSE-06, B3LYP, etc are highly computationally expensive which is beyond the computational resources available to us. Therefore we have further investigated the systems with SCAN meta-GGA functional. The obtained results are consistent with the experiment as well as previously reported simulated results. The contribution of different orbitals at the band edges is evident from the total and orbital projected density of states analysis as given in Fig. 2b-f. From Fig. 2b, it is clear that the valence band  of C$_2$N is dominated by the C `2p' states with a significant contribution from N `2p' states as well. The CBM minimum is also populated by the hybridization of the `p' states of  C and N atoms. The partial DOS analysis of MoS$_2$ establishes that the VBM is highly dominated by Mo `4d' states with a relatively less population than S `3p' states. On the other hand, the CBM is mainly populated by Mo `4d' states. The projected DOS for the heterostructure indicates that the VBM is populated by the states from MoS$_2$ and the CBM is dominated by the C$_2$N. 

\paragraph{}
It has been established that strain plays a significant role in efficiently tuning the electronic, optical, and photocatalytic properties of van der Waals heterojunctions. In this section, we studied the effect of vertical strain by changing the interlayer distance between the C$_2$N and MoS$_2$ monolayers. However, we would like to mention that  experimentally, the interlayer distance can be  effectively varied by changing pressure with a scanning tunneling microscopy tip~\cite{Mudd2016}, through vacuum thermal annealing~\cite{Tongay2014}, inserting a dielectric BN layer inside the van der Waals gap of the heterostructure~\cite{Fang2014}, using diamond anvil cells~\cite{yankowitz2016pressure}. The vertical strain ($\Delta D$) can be defined as $\Delta D=\frac{d-d_0}{d_0} \times 100$, where $d_0$ and $d$ are the interlayer separation between C$_2$N and MoS$_2$ under equilibrium and strained configurations, respectively. The effect of vertical strain on the band gap of the heterostructure is given in Fig. 3a. It is observed that with an increase in tensile strain along the `z' direction, the band gap value shows an increasing trend and exhibits a linear variation. Instead, when the compressive strain between the layer is increased, the band gap decreases and follows a linear relationship with the direction of applied strain.  The band evolution is further analysed through the strain-dependent projected density of states (PDOS) as shown in supporting information S1. It is clear that the compressive strain increases the population of Mo `4d' states for the VBM near the Fermi level. The CBM is mainly composed of a mixture of `p' states of C$_2$N and MoS$_2$. With an increase in tensile strain both Mo `4d' and C$_2$N, MoS$_2$ `p' states move away from the Fermi level resulting in a larger band gap. The band gap calculation is further analysed using the meta-GGA SCAN functional which indicates an enhanced band gap value as compared to the PBE result for the entire range of compressive and tensile strains considered. The band gap variation for the compressive strain shows a nearly linear behaviour consistent with the PBE result. Similarly, for the tensile configuration, the energy gap exhibits an increase in value up to 3\% strain and then remains constant with further expansion in interlayer separation.

\begin{figure}[!htb]
\centering
\includegraphics[width=1.0\textwidth]{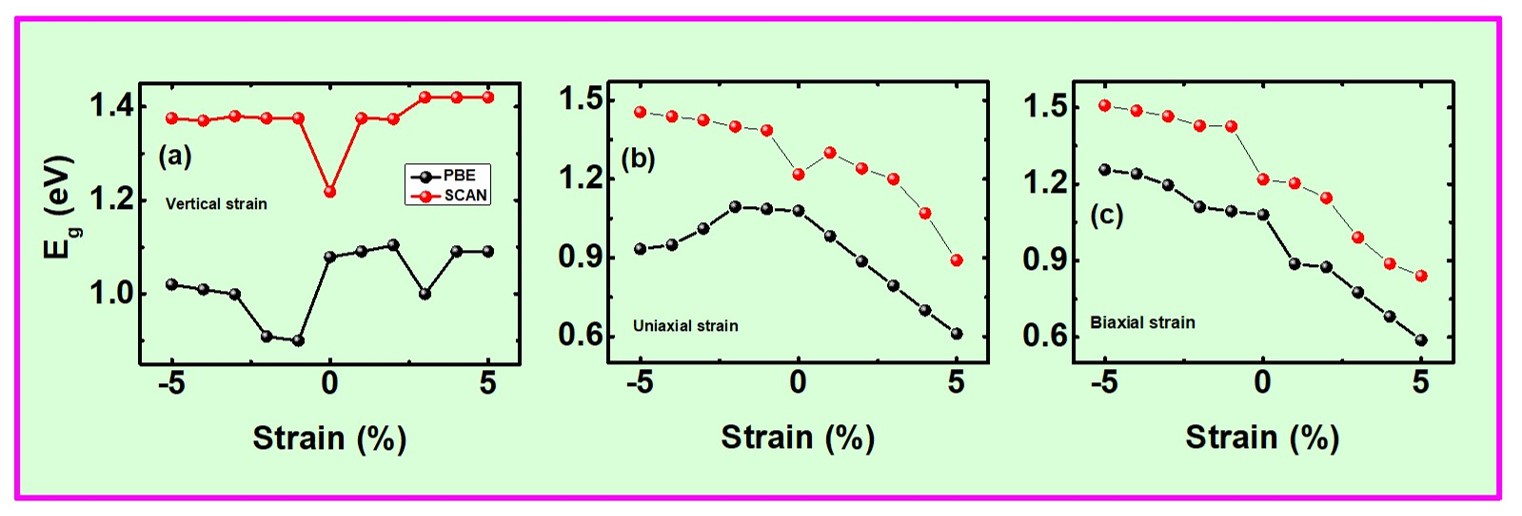}
\caption{Band gap tuning of C$_2$N/MoS$_2$ heterostructure under (a) vertical, (b) uniaxial, and (c) biaxial strain using PBE and SCAN functional }
\label{Schematic}
\end{figure}

The present heterostructure is also investigated with the application of uniaxial and biaxial strain in order to study the influence of in-plane orbital overlapping. The PBE result for the band gap evolution as a function of uniaxial strain is represented in Fig. 3b. It is interesting to note that the band gap remains nearly linear for larger compressive strain. As the compressive strain decreases from -5\% to -1\% the band gap shows a mild increase in value. The application of tensile strain along the uniaxial direction shows a different trend as compared to the compressive one. The band gap shows a systematic decrease in value with the increase in tensile strain from 0 to +5\%. The reduction in band gap may be due to the decrease in orbital overlapping between the `p' states of C and N and the `4d' states of Mo. The corresponding partial density of state (See supporting information Figure S2a, b) suggests that tensile strain increases the DOS near the Fermi level for Mo `4d' and C, N `p' states. Similarly, the band gap study was also executed by applying biaxial strain from -5\% to +5\% range. The system shows a completely different behaviour as compared to the result for vertical strain. The linear variation of the band gap for the compressive strain region remains preserved as similar to the case of uniaxial strain with a linear variation with a slowly decreasing trend. However, if we apply tensile strain then the band gap decreases much more rapidly. The corresponding PDOS analysis indicates that for -5\% compressive strain, the VBM is mainly composed of C and N `2p' states, and CBM is populated  by `p' states of N atoms. When the strain increases from -5\% to +5\%, Mo `4d' states become the highest occupied band near the Fermi level. The CBM comes closer to the Fermi level and resulting in the reduced band gap. The band gap calculation is further analysed using SCAN functional for the uniaxial and biaxial strain configuration which provides an enhanced band gap value as compared to the PBE functional. However, the trend in variation of the band gap remains almost the same (Fig. 3b, c).

The essential requirement for any material for efficient photocatalytic water splitting application is to identify appropriate band edge positions relative to the hydrogen evolution reaction (HER) and oxygen evolution reaction (OER) potentials of the water. The CBM of the material should lie above the reduction reaction potential and the VBM should lie below the oxidation reaction potential of water. To estimate the band edge positions relative to the oxidation and reduction potential, the absolute energy position of the CBM and VBM are calculated with respect to the vacuum level. The vacuum level for the material is obtained by calculating the local potential and then taking the average of the constant potential region. In the present study, we have investigated the photocatalytic activity of the C$_2$N/MoS$_2$ heterostructure by applying strain in the uniaxial, biaxial, and vertical directions. We note that the reduction potential (E$_{H^+}/$H$_2$) and oxidation potential (E$_{O_2}/$H$_2$O) of water with respect to vacuum level are -4.44 and -5.67 eV, respectively~\cite{weast1986handbook}. The band alignment is then calculated by plotting the appropriate band edge position with respect to the vacuum level as a function of strain.

\begin{figure}[!htb]
\centering
\includegraphics[width=1.0\textwidth]{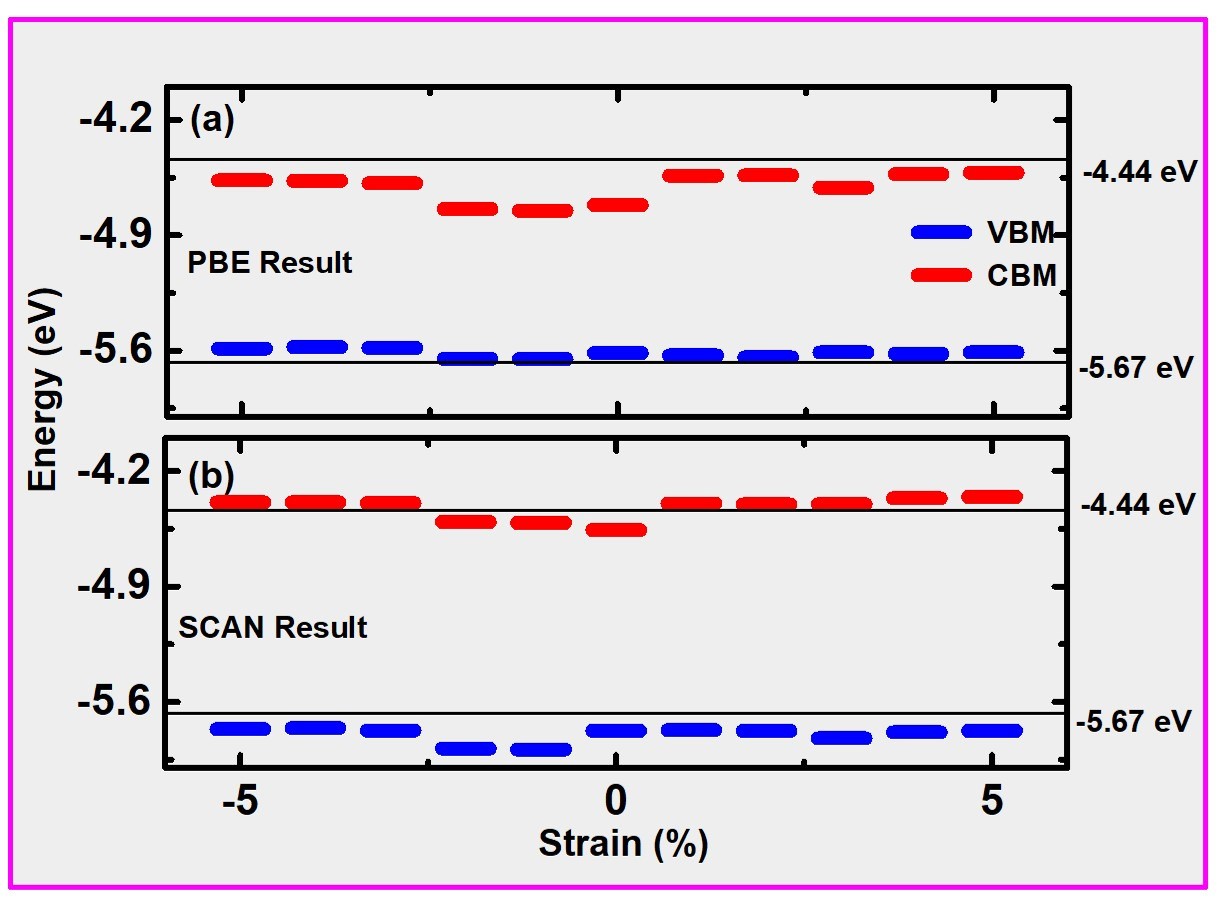}
\caption{Band edge position of C$_2$N/MoS$_2$ heterostructure as a function of vertical strain with respect to vacuum potential using (a) PBE and (b) SCAN functional. The horizontal solid lines represent the water redox potentials }
\label{Schematic}
\end{figure}

The calculated band edge position of the CBM and VBM under the influence of vertical strain, using PBE functional is presented in Fig. 4a. It was observed that without applying any strain, the heterostructure exhibits band edge positions within the water redox potentials of the water. This result indicates that the sample is not useful to carry out photocatalytic water splitting. The sample is subjected to compressive (tensile) vertical strain by reducing (increasing) the interlayer separation between the C$_2$N and MoS$_2$ monolayers, respectively. It was observed that by applying compressive strain, the band edge positions got improved as compared to the zero strain case. However, both the CBM and VBM lie within the water redox potential values. Applying tensile strain, the VBM and CBM move closer to the water redox potential values but still lie below them. Therefore the PBE result indicates that the application of -5 to +5\% vertical strain does not enhance the band edge positions to be useful for photocatalytic water splitting. However, we would like to mention that PBE functional severely underestimates the band gap and band edge positions. Therefore the calculations are further executed using meta-GGA SCAN functional (Fig. 4b).  The result indicates a significant improvement as compared to that of the PBE functional. It was observed that, without applying any strain, the VBM comes below the oxidation reaction potential, whereas the CBM moves up but still lies below the reduction potential value. The application of compressive strain up to -2\% positions the VBM well below the oxidation potential of water but the CBM still lies below the reduction potential. On further increase in the compressive strain from -3 to -5\%, both the CBM and VBM lie outside the range of water redox potentials hence enhancing the photocatalytic activities of the sample. When the heterostructure is subjected to vertical tensile strain from +1\% to +5\%, both the CBM and VBM straddle the water redox potential range hence enhancing the photocatalytic response under visible light. There fore we conclude that the present heterostructure exhibits efficient charge separation for photocatalytic water splitting under tensile strain and larger compressive strain.

The sample is further subjected to uniaxial and biaxial strain to analyse the photocatalytic performance. The uniaxial strain is applied by increasing or compressing the in-plane lattice constant along the X- direction keeping the Y value fixed. Similarly, the biaxial strain is applied by increasing or decreasing the X-Y lattice constant values simultaneously. It is to be noted that the interlayer separation is kept at its equilibrium value i.e 3.373 \AA. The detailed graphical representation of band alignment under uniaxial strain is given in the Supporting information S3 a. It was observed that for PBE functionals the CBM and VBM both lie within the redox potential range for the entire range of uniaxial strain from -5 to +5\%, thus indicating that uniaxial strain does not produce a significant improvement to be useful for photocatalytic water splitting.

To get a better result, the strain calculation is repeated for the heterostructure using SCAN functional and is given in Supporting information S3b. It was observed that the application of larger uniaxial strain (-3 to -5\%) puts the CBM and VBM outside the water redox potential range. But the tensile strain reduces the band edge separation even below the unstrained case. The VBM lies below the oxidation potential value but the CBM finds its position below the reduction potential. Therefore we found that the sample can be useful for photocatalytic water splitting for larger compressive uniaxial strain whereas the tensile strain does not produce an enhancement in the photocatalytic response. The present C$_2$N/MoS$_2$ heterostructure is also investigated under biaxial strain using both PBE and SCAN functionals (See Supporting information S4). The results are quite similar to that of the uniaxial strain. Like previous results, PBE functional does not produce a better result for the photocatalytic activity. A larger compressive strain puts the CBM above the reduction potential but the VBM lies above the oxidation potential of water, thus not providing a useful result for our purpose. Using SCAN functional, it was observed that from -2 to -5\% compressive strain both the CBM and VBM straddle outside the water redox potential. On the other hand, tensile strain decreases the band edges position and put the CBM below the reduction potential of water. Therefore, the heterostructure can be used for photocatalytic studies for larger uniaxial and biaxial compressive strains.

\begin{figure}[!htb]
\centering
\includegraphics[width=1.0\textwidth]{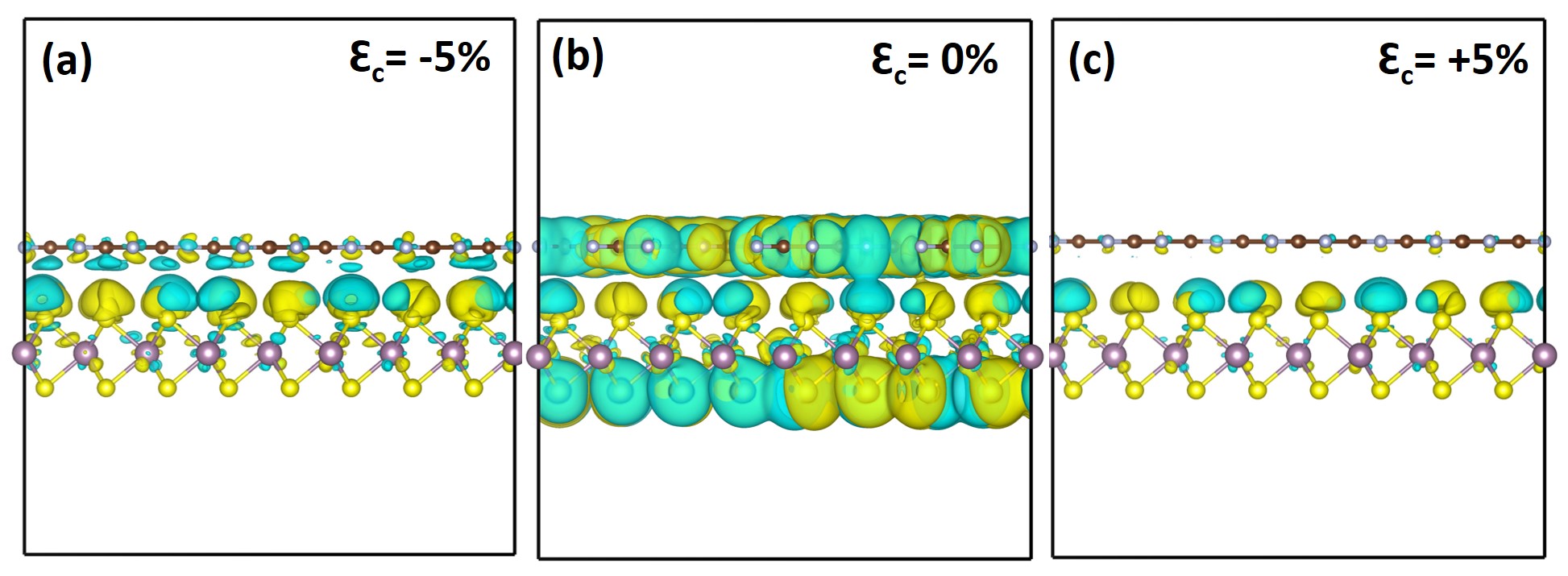}
\caption{Charge density difference of C$_2$N/MoS$_2$ heterostructure as a function of vertical (a) -5\% compressive (b) 0\% (c) +5\% tensile strain}
\label{Schematic}
\end{figure}

In order to illustrate the charge transfer process between the C$_2$N and MoS$_2$ monolayers during the formation of the heterostructure, the charge density difference is executed using PBE functional. The charge density difference is calculated by subtracting the charge density of the individual C$_2$N and MoS$_2$ monolayers from the C$_2$N/MoS$_2$ heterostructure.The charge density difference for the unstrained heterostructure is illustrated in Fig. 5b. In all our charge density figures, the cyan color indicates the charge depletion and the yellow color represents the charge accumulation process. From Fig. 5b, we observe that the charge accumulation mainly occurs in the interface region and partially in  MoS$_2$ monolayers, whereas most of the charge depletion occurs in the top C$_2$N and bottom MoS$_2$. The change in charge density under vertical strain is given in Fig. 5.

From the Fig.5a, we observe that with an increase in compressive vertical strain by 5\%, there is an equal proportion of charge accumulation and depletion close to the MoS$_2$ layer whereas the charge depletion mainly occurs in the region close to C$_2$N layer. The charge density given in cyan color near the C atom in the C$_2$N monolayer indicates a charge depletion. Similarly, the charge density given in yellow color near the S atom in the MoS$_2$ layer indicates a charge accumulation. Hence this observation indicates that there is a charge transfer from the S atom of MoS$_2$ to the C atom of C$_2$N monolayer. The intensification of the charge transfer process under large compressive strain indicates an enhanced interaction between the two monolayers. The strain-induced charge transfer in C$_2$N based heterostructures is consistent with other literature reports~\cite{Guan2017, AshwinKishore2020, Wang_2_2022}. When the interlayer separation in increases to achieve a strain around +5\%, the charge depletion from the C$_2$N layer almost disappears and most of the charge accumulation occurs in the interface region close to MoS$_2$. In other words, there is no appreciable charge redistribution close to the MoS$_2$ layer. This may be the influence of the van der Waals gap between the two monolayers which varies as the structure moves from a compressive strain state to a tensile strain state.

The effect of uniaxial compressive and tensile strain on the charge transfer process is illustrated in supplementary information Fig. S5. It indicates that there is no appreciable change in the charge density difference under uniaxial compressive and tensile strain (up to 5\%). On the other hand application of biaxial strain indicates a significant influence on the charge density under compressive and tensile strains. From the Supplementary information Fig. S6, we observe that as the system is subjected to -5\% compressive strain the charge depletion occurs in the C atom of the top layer. With system changes from large compression to large tensile strain state, the depletion near the C atom increases indicating more charge transfer from the S atom of MoS$_2$ to the C atom of C$_2$N layer. Therefore we observe that both vertical and biaxial strain affects the charge transfer process significantly as compared to the uniaxial strain state.

\section{Conclusion}
   
In summary, we have studied the photocatalytic performance of C$_2$N/MoS$_2$ heterostructure as a function of uniaxial, biaxial, and vertical strain configuration through first-principles DFT calculations. The unstrained heterostructure possesses a direct band gap of 1.35 eV, where the VBM is populated by Mo `d' states and the CBM is contributed by `C' and `N' `p' states. The calculated position of CBM and VBM of individual monolayers  indicates that the heterostructure possesses a type-II band alignment which is beneficial for charge separation across the two monolayers and preventing their recombination process. The SCAN functional provides better results for the band gap and band edge positions calculation  as compared to that of the PBE result. Using the C$_2$N/MoS$_2$ heterostructure as a prototype, we have also found that the meta-GGA SCAN functional shows similar results as compared to the computationally expensive hybrid HSE functionals. The estimated CBM and CBM position as a function of vertical strain indicates that the water reduction and oxidation potential values lie within the band gap region with respect to the vacuum level for larger compressive and tensile strain. In contrast, the system exhibits good photocatalytic performance only for larger compression for uniaxial and biaxial strain states whereas the tensile strain reduces the separation between the VBM and CBM within the water redox potential value. The charge density difference indicates a significant charge transfer for vertical and biaxial configuration as compared to the uniaxial state. The present study can help researchers to reduce the computational cost by considering the meta-GGA functionals over HSE for electronic structure calculations of similar systems. Our calculation will also be extremely useful for designing artificial strained heterostructure for the experimental community for better device application for photocatalytic water splitting. 

\section{Computational Methods}
 The first principles electronic structure calculations were performed using density functional theory (DFT) with projector augmented-wave method~\cite{blochl1994projector} as implemented in the Vienna ab initio simulation package (VASP)~\cite{kresse1996efficient}. The Perdew-Burke-Ernzerhof (PBE)~\cite{perdew1996generalized} parametrization-based generalized gradient approximation (GGA) was chosen for the  exchange-correlation functional. To accurately describe the interaction between the layered structures, we have included the van der Waals correction method (DFT-D2) presented by Grimme~\cite{Grimme2006}. Initially, the gap between C$_2$N  and MoS$_2$ monolayers is kept at 3.5~\AA ~which is further optimized before running the electronic structure calculation. As a benchmark, the system is further studied using strongly constrained and appropriately normed (SCAN) meta-GGA functionals to get more accurate results as compared to the PBE functional and also to be consistent with the reported experimental data. A vacuum layer of 20~\AA ~was selected along the `Z' direction to avoid interaction between the adjacent layers. The plane wave expansion cut-off was chosen to be 520 eV. The Brillouin zone integration was performed using a $\Gamma$-centered (9$\times$9$\times$1) \textbf{k}-mesh for the structural relaxation and electronic structure calculation of the isolated C$_2$N and MoS$_2$ monolayers. For the  C$_2$N/MoS$_2$ heterostructure containing 354 atoms, the geometry optimization and electronic structure calculation are performed using a $\Gamma$ centered (2$\times$2$\times$1) \textbf{k}-point samplings. All the structures are allowed for relaxation to get the optimized atomic position until the total forces acting on each atom are less than 0.02 eV/\AA.
\section*{Data availability statement}

The additional data that support the findings of this article are available in the Supplementary Information.
    
\section*{Acknowledgments}
 The authors would like to thank the National Institute of Science Education and Research (NISER), Department of Atomic Energy, Government of India, for funding the research work through project number RIN-4001. The authors acknowledge the high-performance computing facility at NISER.

\section*{Data availability statement}

The additional data that support the findings of this article are available in the Supplementary Information.
\section*{Conflict of interest}
The authors have no conflicts to disclose.

\bibliography{output}
\end{document}